\documentstyle[preprint,aps,epsfig]{revtex}
\begin{document}
\draft
\title{
Li intercalation effects on magnetism in undoped and Co-doped anatase TiO$_2$}
\author{Min Sik Park, S. K. Kwon, and B. I. Min}
\address{Department of Physics,
        Pohang University of Science and Technology, 
        Pohang 790-784, Korea}
\date{\today}
\maketitle
\begin{abstract}
The effects of $n$-type carrier doping by Li intercalation on magnetism in 
undoped and Co-doped anatase TiO$_2$ are investigated.
We have found that doped $n$-type carriers in TiO$_2$ are localized mainly 
at Ti sites near the intercalated Li.
With increasing the intercalation, local spins are realized at Ti.
In the case of Co-doped TiO$_2$, most of the added 
$n$-type carriers fill the Co 3$d$ bands and 
the rest are localized at Ti.
Therefore, Co magnetic moment vanishes by Li intercalation to have
a nonmagnetic ground state. 

\end{abstract}
\pacs{Keywords: Dilute magnetic semiconductor; Electronic structure; TiO$_2$ 
\\
}

Anatase TiO$_2$ is a wide band gap (3.2 - 3.7 eV) semiconductor.
This wide band gap property provides a wide range of 
applications, such as air and water purifications using photocatalysis 
which converts solar energy into electrochemical energy\cite{Asahi}
and batteries, 
electrochromic devices based on lithium intercalation.
A theoretical study\cite{Stashans} shows that Li 
intercalation is easier in anatase TiO$_2$ than in rutile TiO$_2$.
This is due to the open structure of anatase TiO$_2$.
Recently, magnetic features are observed in the Li intercalated
anatase TiO$_2$\cite{Luca}. 
It is possible to intercalate Li atoms up to Li/Ti ratio $\sim$ 0.7.
With increasing the Li intercalation,
a structural transition occurs from tetragonal to orthorhombic one.
Also the insulator-to-metal transition is observed for $x$$>$ 0.3.
The localized moments were measured, 0.003 $\sim$ 0.004$\mu_B$
per Ti, through the Li intercalation. 

On the other hand,  room temperature ferromagnetism 
is observed in Co-doped anatase TiO$_2$ thin film made by the combinatorial
pulsed-laser-deposition molecular-beam epitaxy method\cite{Matsumoto}.
A sizable amount of Co, up to 8$\%$, is soluble in anatase TiO$_2$.
The measured saturated magnetic moment per Co ion was 0.32$\mu_B$ 
apparently in the low spin state and $T_C$ was estimated to be 
higher than 400 K. 
It is recognized that the carriers play role of inducing the ferromagnetism 
in dilute magnetic semiconductors\cite{Dietl}.

To explore the carrier doping effects on magnetism,
we have investigated electronic and magnetic properties of Li intercalated
anatase TiO$_2$ and Ti$_{1-x}$Co$_x$O$_2$ ($x$=0.0625).
Li/Ti intercalation ratios of 0.0625, 0.125, and 0.25 are considered for 
the undoped case, while 0.067 and 0.133 for the Co-doped case. 
We consider only the tetragonal anatase structure.
We have used the linearized muffin-tin orbital (LMTO) band method in the
local-spin-density approximation (LSDA). 
We have considered a supercell containing 16 f.u. in the primitive unit cell
by replacing one Ti by Co or intercalating several Li ions
($a$=$b$=7.570, $c$=9.514 $\rm \AA$). 
Sixteen empty spheres are employed at the interstitial sites to
enhance the packing ratio in the LMTO band calculation.

We have first calculated the electronic structure of Li intercalated
TiO$_2$. In all Li/Ti ratio cases, we have obtained metallic ground 
states (Fig.~\ref{undoped}).
For Li/Ti=0.0625, the paramagnetic ground state is obtained. 
However, in other cases, total energies of the paramagnetic 
and the ferromagnetic ground state are almost the same.
Doped $n$-type carriers fill the Ti 3$d$ conduction band.
Maximum localized magnetic moments at Ti are 0.029, 0.027 $\mu_B$
for Li/Ti=0.125 and 0.25, respectively.
Thus the magnetic moment is not necessarily
proportional to the number of localized electrons at Ti.
The exchange splitting is clearly seen in the valence band top mainly 
with O 2$p$ characters (Fig.~\ref{undoped}).

Now, we have performed band calculations for Co-doped case. For Li/Ti=0.067,
we have obtained paramagnetic and insulating ground state 
in Fig.~\ref{doped}. 
Insulating ground state results from filling up Co 3$d$ band
by the $n$-type carriers. 
Therefore, Co $t_{2g}$ band is fully occupied, and the 
total occupancy of $d$ states amounts to $d^7$($t_{2g}^6e_g^1$). 
As seen in Fig.~\ref{doped}, the position of occupied $t_{2g}$ band 
is different from the non-intercalated Co-doped case. 
In the latter case, the occupied Co $t_{2g}$ band is located 
near the valence band top\cite{Mspark},
while, in the former case, the fully occupied Co $t_{2g}$ band
is located near the conduction band bottom.
The unoccupied Co $e_g$ band is hybridized with Ti 3$d$ conduction band. 

For the Li/Ti=0.133 case, nearly paramagnetic and metallic ground state is
obtained (Fig.~\ref{doped}). 
The carriers are mainly of Ti 3$d$ character.
The extra electrons after filling up the Co $t_{2g}$ band occupy
not Co $e_g$ band but Ti 3$d$ conduction band, because the carriers 
are localized at Ti sites  near the intercalated Li.
From these, one can expect that $n$-type carrier induced
ferromagnet can be fabricated in TiO$_2$ by simultaneously intercalating Li 
and doping some 3$d$ transition metals 
having high spin magnetic ground state such as Mn, Fe\cite{Mspark}.

In conclusion,  
we have found that, by intercalating Li in TiO$_2$, Ti atoms have
localized magnetic moments, 0.029 and 0.027$\mu_B$
for Li/Ti=0.125 and 0.25, respectively. 
In the case of Co-doped TiO$_2$, nonmagnetic ground states
are obtained for Li/Ti=0.067 and 0.133.

Acknowledgments$-$
This work was supported by the KOSEF through the eSSC at POSTECH
and in part by the BK21 Project.


\begin{figure}[t]
\caption{The LSDA total and projected local density of states (PLDOS) 
         of Li intercalated TiO$_2$ ((a),(b) : Li/Ti=0.0625, 
         (c),(d) : Li/Ti=0.125)} 
\label{undoped}
\end{figure}

\begin{figure}[t]
\caption{The LSDA total and PLDOS of Li intercalated $\rm Ti_{1-x}Co_xO_2$
         ($x$=0.0625) ((a),(b) : Li/Ti=0.067, (c),(d) : Li/Ti=0.133)}
\label{doped}
\end{figure}

\newpage
\centerline{Fig.1}
\centerline{\epsfig{figure=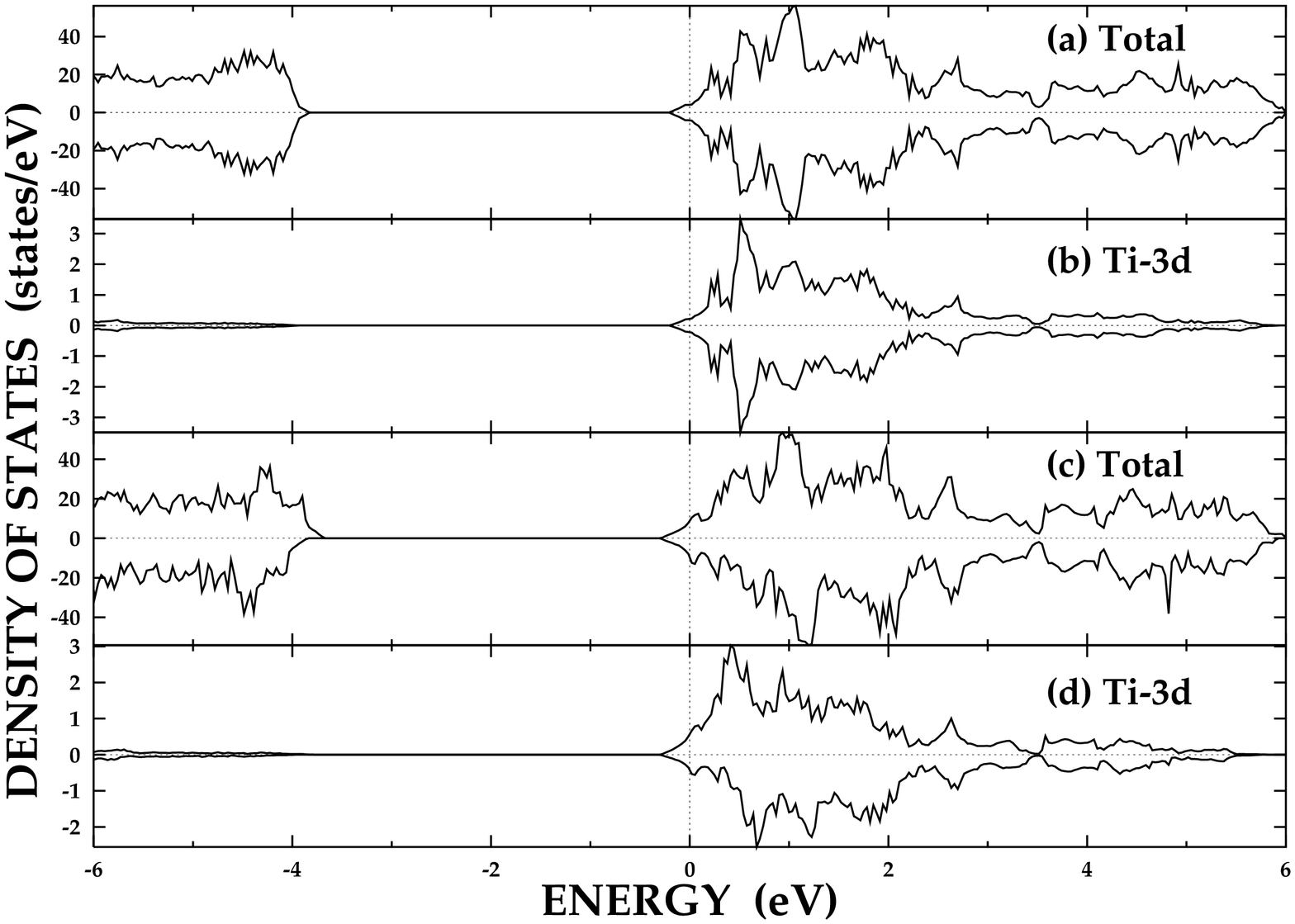,angle=270,width=12cm}}

\centerline{Fig.2}
\centerline{\epsfig{figure=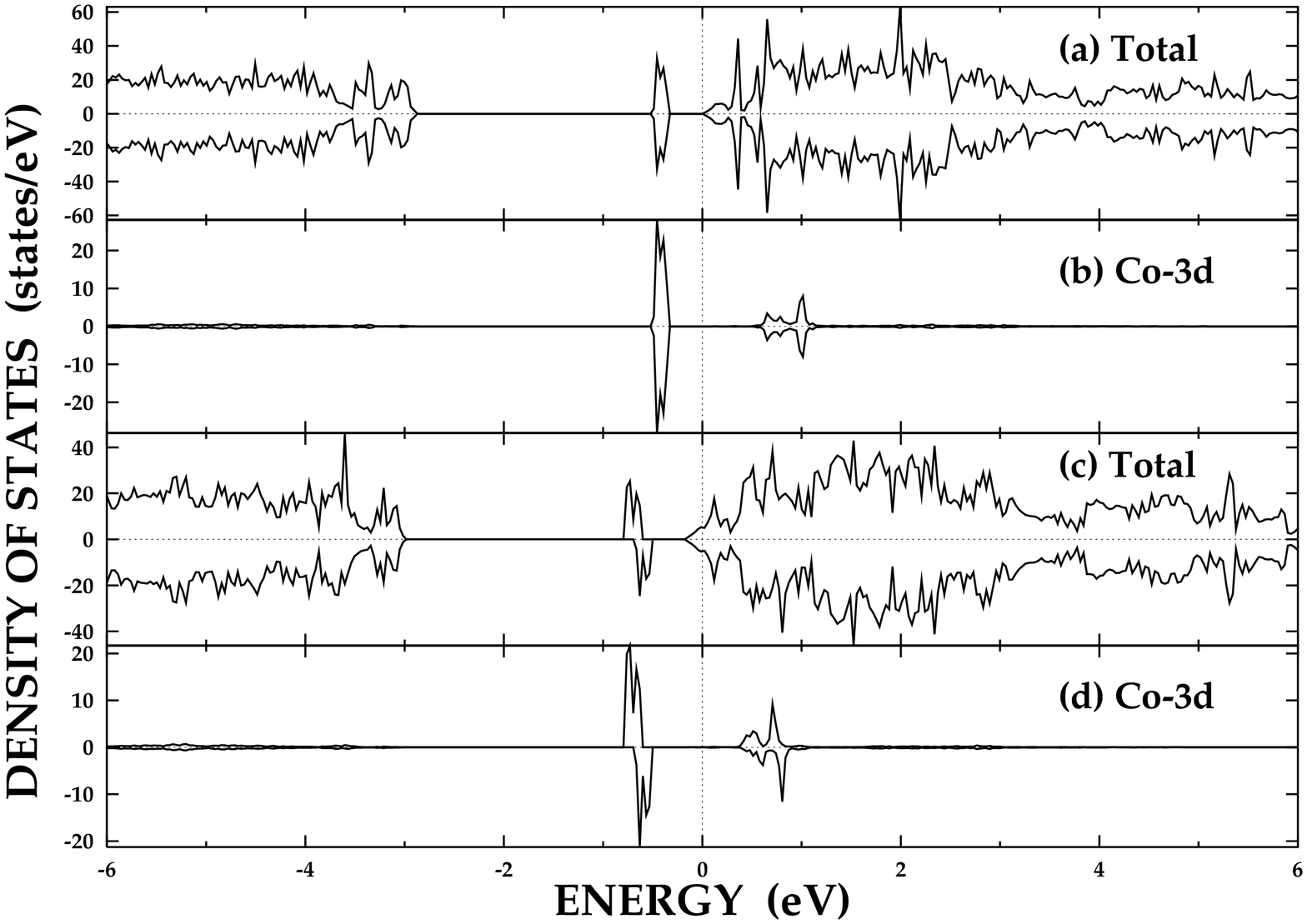,angle=270,width=12cm}}


\begin{references}
\bibitem{Asahi} R. Asahi {\it et al.}, Science 293 (2001) 269.
\bibitem{Stashans} A. Stashans {\it et al.}, Phys. Rev. B 53 (1996) 159.
\bibitem{Luca} V. Luca {\it et al.}, Chem. Mater. 13 (2001) 796.
\bibitem{Matsumoto}  Y. Matsumoto {\it et al.}, Science 291 (2001) 854.
\bibitem{Dietl} T. Dietl {\it et al.}, Science 287 (2000) 1019.
\bibitem{Mspark} M. S. Park {\it et al.}, Phys. Rev. B 65 (2002) 161201.


\end{references}
\end{document}